\documentclass{article}
\usepackage{amsmath}
\usepackage{amsfonts}
\usepackage{amssymb}
\usepackage[usenames]{color}
\usepackage{amsthm}
\def\bbox#1{{\mathbf{#1}}}
\def\Th{\Theta}
\def\l{\lambda}
\def\p{\partial}
\def\e{\mathrm{e}}
\newtheorem{prop}{Proposition}

\theoremstyle{remark}

\newcommand{\dbar}{\bar{\partial}}
\newcommand{\wt}{\widetilde}
\newcommand{\be}{\begin{equation}}
\newcommand{\ee}{\end{equation}}
\newcommand{\bea}{\begin{eqnarray}}
\newcommand{\eea}{\end{eqnarray}}
\newcommand{\beaa}{\begin{eqnarray*}}
\newcommand{\eeaa}{\end{eqnarray*}}

\newcommand{\nn}{\nonumber}
\renewcommand{\d}{\mathrm{d}}

\usepackage{authblk}
\begin{document}
\title
{Matrix extension of multidimensional dispersionless
integrable hierarchies
\thanks
{This research was performed in the framework 
of State Assignment of Ministry of Science and
Higher Education of the Russian Federation,
topic 0029-2021-0004 
(Quantum field theory).}
}
\author{L.V. Bogdanov \thanks{leonid@itp.ac.ru}}
\affil{Landau Institute for Theoretical Physics RAS}
\date{}
\maketitle
\begin{abstract}
We consistently develop a recently proposed scheme
of matrix extension of dispersionless integrable systems
for the general case of  multidimensional hierarchies,
concentrating on the case 
of dimension $d\geqslant 4$. We present extended
Lax pairs, Lax-Sato equations, matrix equations
on the background of vector fields
and the dressing scheme. Reductions,
construction of solutions and connections to geometry
are discussed.
We consider separately a case of Abelian
extension, for which the Riemann-Hilbert equations of
the dressing scheme are explicitly solvable and give an
analogue of Penrose formula in the curved space.
\end{abstract}

\section{Introduction}
Multidimensional dispersionless integrable systems 
correspond to the Lax pairs of the type
\bea
&&
[X_1, X_2]=0,
\nn
\\&&
X_1=\p_{t_1}
+\sum_{i=1}^N F_i\p_{x_i} + F_0\p_\lambda,\quad
X_2=\p_{t_2}+ \sum_{i=1}^N G_i\p_{x_i}+G_0\p_\lambda.
\label{Lax}
\eea
where $\lambda$ is a `spectral parameter', functions $F_k$, 
$G_k$ are holomorphic in $\lambda$ and depend 
on the variables $t_1$, $t_2$, $x_n$. 
We will consider polynomials
(or Laurent polynomials) in $\lambda$.
This class includes
dispersionless limits of integrable 
(2+1)-dimensional equations 
(dispersionless Kadomtsev-Petviashvili equation, 
dispersionless 2DTL hierarchy) and equations of
complex relativity, integrable by the twistor method 
(Pleba\'nski heavenly equations, 
hyper-K\"ahler hierarchies etc.). 

Recently we proposed \cite{LVB17,LVB19,LVB20}
a scheme of matrix extension of Lax pairs (\ref{Lax}),
leading to gauge covariant Lax pairs of the type
\bea
\label{Laxext}
\nabla_{X_1}=X_1 + A_1,
\quad
\nabla_{X_2}=X_2 + A_2,
\eea
$A_1$, $A_2$ are matrix-valued functions of 
space-time variables holomorphic in $\lambda$ 
(polynomials, Laurent polynomials).

Lax pairs of this structure 
were already present 
in the seminal work of
Zakharov and Shabat \cite{ZS} (1979), where it was noticed
that the commutation relation splits into (scalar) vector
field part
and Lie algebraic part, which gives the equations similar
to the self-dual Yang-Mills equations 
but on some background the origin of which 
was not clear at the moment.

Indeed, the commutator of two covariant vector fields
is itself of the form of covariant vector field, it 
contains
vector field part and matrix (Lie algebraic) part,
\beaa
[\nabla_{X_1}, \nabla_{X_2}]=
[X_1, X_2]+ X_1 A_2 - X_2 A_1 +[A_1,A_2]
\eeaa
Compatibility condition contains a vector fields part 
($(N+2)$-dimensional dispersionless system)
\beaa
[X_1, X_2]=0
\eeaa
and a matrix part (matrix equations on the 
dispersionless background)
\beaa
X_1 A_2 - X_2 A_1 +[A_1,A_2]=0
\eeaa
To define a closed system of matrix equations for the 
coefficients of polynomials (meromorphic functions)
$A_1$, $A_2$, the holomorphic structure of $A_1$, $A_2$
(the order of zeroes and poles) should be consistent
with the holomorphic structure of vector fields,
and the gauge should be fixed (it is also possible
to consider gauge-invariant matrix equations without fixing the
gauge). The scheme of matrix extension we discuss below 
allows to construct prolongation terms $A_1$, $A_2$ with
a correct structure, satisfying a compatibility condition,
thus giving a solution to matrix equations 
on the dispersionless background.

Compatibility conditions for the Lax pair (\ref{Lax})
imply (local) existence of common solutions 
for linear equations
\beaa
X_1\psi_i=0,\quad X_2\psi_i=0.
\eeaa
These equations, according to Frobenius theorem,
have $N+1$ functionally independent solutions 
$\psi_i(\lambda,t_1,t_2, \mathbf{x})$  -- `wave fuctions' 
of linear operators of background
dispersionless integrable system, a general solution is 
\beaa
\psi=f(\psi_0,\dots,\psi_N)
\eeaa
Let us introduce a matrix-valued
wave function for the extended system
\beaa
\nabla_{X_1}\Phi_0=(X_1 + A_1)\Phi_0=0,
\quad
\nabla_{X_2}\Phi_0=(X_2 + A_2)\Phi_0=0.
\eeaa
Locally $\Phi$ may be considered as series in $\lambda$
with matrix coefficients.
A general solution to linear equations is of the form
\beaa
\Phi=\Phi_0 F(\psi_0,\dots,\psi_N),
\eeaa
where $F$ is a matrix-valued complex analytic function of
$(N+1)$ variables.
It is easy to check that extended linear equations 
are equivalent to
\beaa
(X_1 \Phi)\Phi^{-1}=-A_1, \quad
(X_2 \Phi)\Phi^{-1}=-A_2
\eeaa
where $A_1$, $A_2$ are polynomials in $\lambda$ 
with matrix coefficients (Laurent polynomials,
meromorphic functions). 
This is a characteristic analytic property
of $\Phi$, important for algebraic definition of 
the hierarchy (Lax-Sato equations) and for 
construction of solutions by means of Riemann-Hilbert
problem.

To construct $\Phi$, we may consider matrix  
Riemann-Hilbert problem
\beaa
\Phi_{\text{in}}=\Phi_{\text{out}}R(\psi_1,\dots,\psi_N),
\eeaa
defined on some curve  in the complex plane
(we will usually consider the unit circle),
where $\psi_i(\lambda,\mathbf{t})$ are wave functions of 
the dispersionless Lax pair. 
This problem provides analyticity
of the functions $(X_1 \Phi)\Phi^{-1}$, 
$(X_2\Phi)\Phi^{-1}$, thus leading to extended Lax pair
(\ref{Laxext}) and defining the matrix prolongation
functions $A_1$, $A_2$.

In the present work we consistently develop
a scheme of matrix extension of dispersionless
integrable systems for the general case of 
multidimensional hierarchies with polynomial
in spectral variable vector fields,
mainly for dimension $d\geqslant 4$.
We briefly describe the picture of the basic
dispersionless hierarchy. On the background of 
the basic hierarchy we construct the matrix
extension, connected with the introduction of
gauge covariant vector fields.
We present extended
Lax pairs, Lax-Sato equations, matrix equations
on the background of vector fields
and the dressing scheme. Reductions
and construction of solutions 
are discussed.
We consider separately a case of Abelian
extension, for which the Riemann-Hilbert problem
equations of
the dressing scheme are explicitly solvable and give an
analogue of Penrose formula in the curved space.

\section{Basic hierarchy}
A
general (N+2)-dimensional hierarchy 
with polynomial vector fields \cite{BDM07}, \cite{LVB09}
is defined by the generating relation
\bea
(\Omega_0)_-:=(J_0^{-1}\d \Psi^0\wedge \d \Psi^1\wedge \dots \wedge \d \Psi^N)_-=0,
\label{analyticity0}
\eea
where $\Psi^0,\dots, \Psi^N$ are the series
\bea
&&
\Psi^0=\lambda+\sum_{n=1}^\infty \Psi^0_n(\mathbf{t}^1,\dots,\mathbf{t}^N)\l^{-n},
\label{form0}
\\&&
\Psi^k=\sum_{n=0}^\infty t^k_n (\Psi^0)^{n}+
\sum_{n=1}^\infty \Psi^k_n(\mathbf{t}^1,\dots,\mathbf{t}^N)(\Psi^0)^{-n},
\label{formk}
\eea
$1\leqslant k\leqslant N$, $\mathbf{t}^k=(t^k_0,\dots,t^k_n,\dots)$,
$(\cdots)_-$ is a projection to negative powers of $\lambda$,
$J_0$ is a determinant of Jacobian matrix $J$,
\bea
J_0=\det J, \quad J_{ij}=\partial_i \Psi^j,\quad 0\leqslant i,j \leqslant N,
\label{J_0}
\eea
where $\partial_0=\frac{\partial}{\partial \lambda}$,
$\partial_k=\frac{\partial}{\partial x^k}$ ($1\leqslant k \leqslant N$),
$x^k=t^k_0$.

Generating relation (\ref{analyticity0}) is equivalent to the set
of Lax-Sato equations
\bea
&&
\partial^k_n\mathbf{\Psi}=\sum_{i=0}^N\left((J^{-1})_{ki} (\Psi^0)^n)\right)_+
{\partial_i}\mathbf{\Psi},\quad 0\leqslant n\leqslant \infty\, ,
1\leqslant k \leqslant N,
\label{genSato}
\eea
where $\mathbf{\Psi}=(\Psi^0,\dots,\Psi^N)$,
$(\cdots)_+$ is a projection to nonnegative powers of $\lambda$.
First flows of the hierarchy read
\bea
\partial^k_1\mathbf{\Psi}=(\lambda \partial_k-\sum_{p=1}^N (\partial_k u_p)\partial_p-
(\partial_k u_0)\partial_\lambda)\mathbf{\Psi},\quad 0<k\leqslant N,
\label{genlinear}
\eea
where $u_0=\Psi^0_1$,
$u_k=\Psi^k_1$, $1\leqslant k\leqslant N$.
A compatibility condition for any pair of linear equations  
(\ref{genlinear})
(e.g., with $\partial^k_1$ and $\partial^q_1$, $k\neq q$)
implies closed nonlinear 
(N+2)-dimensional  system of PDEs for the set of functions $u_k$, $u_0$,
which can be written in the form
\bea
&&
\partial^k_1\p_q\hat u-\partial^q_1\p_k\hat u+[\p_k \hat u,\p_q \hat u]=
(\p_k u_0)\p_q-(\p_q u_0)\p_k,
\nn\\
&&
\partial^k_1\p_q u_0 - \partial^q_1\p_k u_0 + (\p_k \hat u)\p_q u_0 -
(\p_q \hat u)\p_k u_0=0,
\label{Gensystem}
\eea
where $\hat u$ is a vector field, $\hat u=\sum_{p=1}^N u_p \p_p$. 

In the case $N=2$ we have two vector fields 
(\ref{genlinear}) \cite{BDM07}
\bea
\nabla^1_1
=\partial^1_1 
-\lambda \partial_1
+ (\partial_1 u_1)\p_1
+ (\partial_1 u_2)\p_2
+ (\partial_1 u_0)\partial_\lambda
,
\nn
\\
\nabla^2_1
=\partial^2_1 
-\lambda \partial_2
+ (\partial_2 u_1)\p_1
+ (\partial_2 u_2)\p_2
+ (\partial_2 u_0)\partial_\lambda,
\label{genlinear22}
\eea
giving a Lax pair for a 4-dimensional
closed set of second order
equations for three functions $u_0$, $u_1$, $u_2$
with independent variables $t^1_1$, $t^2_1$, $x^1$, $x^2$,
\begin{gather}
Q u_2=\p_2 u_0,\quad Q u_1=-\p_1 u_0,
\nn
\\
\left(
\partial^1_1 +(\p_1 u_1)\p_1+(\p_1 u_2)\p_2 
\right)
\p_2 u_0 
= 
\left(
\partial^2_1 +(\p_2 u_1)\p_1+(\p_2 u_2)\p_2
\right)
\p_1 u_0  
,
\label{Gensystem2}
\end{gather}
where linear operator $Q$ is
\bea
Q=
\partial^2_1\p_1
-\partial^1_1\p_2 
+(\p_2 u_1)\p_1\p_1 
-(\p_1 u_2)\p_2\p_2 
-((\p_1 u_1)-(\p_2 u_2))\p_1\p_2.
\label{Q}
\eea
This system can be easily rewritten in equivalent form of
two third order equations for functions $u_1$, $u_2$
used in the work \cite{DFK15}, where it was demonstrated 
that it describes a general local form of the self-dual conformal
structure (complex analytic case or real case 
with neutral signature, modulo coordinate transformations).
Linear operator 
$Q$ represents the conformal structure in 
the form of a symmetric bivector (inverse metric),
respectively the metric (representative of the
conformal structure) is given by 
\bea
\label{ASDmetric0}
&&
g=dt^2_1 dx^1
-dt^1_1 dx^2
\nn\\
&&\qquad
-(\p_2 u_1) dt^2_1 dt^2_1
-((\p_1 u_1)-(\p_2 u_2)) dt^2_1 d t^1_1 
+ (\p_1 u_2)dt^1_1 dt^1_1.
\eea

The reduction of the hierarchy to volume-preserving
flows $J_0=1$ corresponds to divergence-free vector fields,
for vector fields (\ref{genlinear22}) it leads to 
introduction of the potential
$\Theta$, $u_1=\Theta_y$, $u_2=-\Theta_x$, 
$x=x^1$, $y=x^2$. After the identification
$z=-t^1_1$, $w=t^2_1$, $\phi=u_0$ we get 
the Dunajski system generalising the second 
heavenly equation \cite{Dun02}
\bea
&&
\Th_{wx}+\Th_{zy}+\Th_{xx}\Th_{yy}-\Th_{xy}^2=\phi,
\nn\\
&&
\label{Dun}
\phi_{xw}+\phi_{yz}+
\Th_{yy}\phi_{xx}+\Th_{xx}\phi_{yy}-2\Th_{xy}\phi_{xy}=0,
\eea
and further reduction to linearly degenerate case
$\phi=0$ gives the second heavenly equation \cite{Pleb}
\bea
\Th_{wx}+\Th_{zy}+\Th_{xx}\Th_{yy}-\Th_{xy}^2=0.
\label{Heav}
\eea
Operator $Q$ defining the conformal structure
is of the same form
for the Dunajski system and
the second heavenly equation,
\beaa
Q=
\partial_w\p_x
+\partial_z\p_y 
+\Theta_{yy}\p_x\p_x
+\Theta_{xx}\p_y\p_y 
-2\Theta_{xy}\p_x\p_y, 
\eeaa
it coincides with the linearisation 
of the second heavenly equation.
Respectively, the metric is given
by the expression
\bea
\label{ASDmetric1}
g=dwdx+dzdy-\Theta_{yy} dw^2 + 2\Theta_{xy}dwdz  
- \Theta_{xx} dz^2.
\eea
\subsection*{Dressing scheme}
The dressing scheme for the hierarchy 
(\ref{analyticity0}),(\ref{genSato})
can be formulated
in terms of (N+1)-component nonlinear 
Riemann-Hilbert problem on the unit circle $S$
in the complex plane of the variable $\l$,
\bea
&&
\Psi^0_\text{in}=F_0(\Psi^0_\text{out},\Psi^1_\text{out},\dots,\Psi^N_\text{out}),
\nn\\
&&
\Psi^k_\text{in}=F_k(\Psi^0_\text{out},\Psi^1_\text{out},\dots,\Psi^N_\text{out}),
\quad 1\leqslant k\leqslant N,
\label{RiemannMS}
\eea
where the functions 
$\Psi^0_\text{in}(\l,\mathbf{t})$, $\Psi^k_\text{in}(\l,\mathbf{t})$ 
are analytic inside the unit circle,
the functions $\Psi^0_\text{out}(\l,\mathbf{t})$, $\Psi^k_\text{out}(p,\mathbf{t})$ 
are analytic outside the
unit circle  and have an expansion of the form (\ref{form0}), (\ref{formk}).
The functions $F_0$, $F_k$ are suggested to define  a
diffeomorphism in $\mathbb{C}^{N+1}$, 
$\mathbf{F}\in\text{Diff(N+1)}$, they represent
the dressing data. 
In compact form the problem 
(\ref{RiemannMS}) can be written as
\be
\mathbf{\Psi}_\text{in}=\mathbf{F}(\mathbf{\Psi}_\text{out}).
\label{RiemannMSbis}
\ee
It is straightforward to demonstrate that the problem
(\ref{RiemannMS}) implies the analyticity of the differential form
\bea
\Omega_0=J_0^{-1}\d \Psi^0\wedge \d \Psi^1\wedge \dots \wedge \d \Psi^N,
\label{omega0}
\eea
(where the independent variables of the
differential include all the times $\mathbf{t}$ and 
the spectral variable $\l$)
in the complex plane and the 
generating relation (\ref{analyticity0}), 
thus defining
a solution of the hierarchy. Considering a reduction to the group of
volume-preserving
diffeomorphisms $\mathbf{F}\in\text{SDiff(N+1)}$, 
we obtain a reduction of the general
hierarchy (\ref{analyticity0}) to the case $J_0=1$ 
(divergence-free vector fields), 
$$
(\d \Psi^0\wedge \d \Psi^1\wedge \dots \wedge \d \Psi^N)_-=0.
$$
\section{Matrix extension}
\label{Ext}
To construct matrix extension of involutive
distribution of
polynomial in spectral parameter vector fields with
the basis $X_i$ we introduce a matrix-valued 
function $\Phi$ possessing the 
property that all the functions $(X_i\Phi)\Phi^{-1}$ are 
polynomial in spectral variable. Existence of 
such functions in terms of series is implied by the 
compatibility of Lax-Sato equations, they can be 
constructed analytically using the Riemann-Hilbert problem
(see below).
The function $\Phi$
is suggested to be bounded and invertible in the
spectral plane and analytic in some neighborhood
of infinity.
Extended linear problems are given by the relations
\bea
X_i\Phi=((X_i\Phi)\Phi^{-1})_+\Phi
\label{ext0}
\eea
where $((X_i\Phi)\Phi^{-1})_+$ is a polynomial part
of the Laurent series at infinity
containing finite number of matrix fields,
and extended vector fields are
\beaa
\nabla_{X_i}=X_i - ((X_i\Phi)\Phi^{-1})_+.
\eeaa
Let us consider canonically normalised $\Phi$ as 
series
\bea
\Phi=
1+\sum_{n=1}^\infty \Phi_n(\mathbf{t})\lambda^{-n}
\label{Phiser0}
\eea
For polynomial vector fields of the form (\ref{Lax}),
relations (\ref{ext0}) are 
\bea
&&
\p_{t_1}\Phi= (-F + ((F\Phi)\Phi^{-1})_+)\Phi,
\nn
\\
&&
\p_{t_2}\Phi= (-G + ((G\Phi)\Phi^{-1})_+)\Phi,
\label{LS00}
\eea
where
\beaa
F=
\sum_{i=1}^N F_i\p_{x_i} + F_0\p_\lambda,\quad
G=\sum_{i=1}^N G_i\p_{x_i}+G_0\p_\lambda.
\eeaa
These relations may be considered as Lax-Sato equations
defining the evolution of the series $\Phi$ with
matrix coefficients depending on the variables
$x_1,\dots,x_N$ with respect to the times
$t_1$, $t_2$.
\begin{prop}
\label{compatibility}
For commuting polynomial vector fields 
$X_1$, $X_2$ of the form (\ref{Lax}), 
Lax-Sato equations (\ref{LS00})
for series (\ref{Phiser0}) 
are compatible.
\end{prop}
\begin{proof}
Compatibility means that cross-derivatives over times
are equal by virtue of equations (\ref{LS00}),
\beaa
\Delta=
\p_{t_1}(-G + ((G\Phi)\Phi^{-1})_+)\Phi
-
\p_{t_2}(-F + ((F\Phi)\Phi^{-1})_+)\Phi
=0.
\eeaa
On the one
hand,
\beaa
\Delta=\left\{
\p_{t_1}((G\Phi)\Phi^{-1})_+
-
\p_{t_2}((F\Phi)\Phi^{-1})_+
+
[((G\Phi)\Phi^{-1})_+,((F\Phi)\Phi^{-1})_+]
\right\}\Phi
\eeaa
and $\Delta \Phi^{-1}$ contains only non-negative powers
in $\lambda$,
$
(\Delta \Phi^{-1})_-=0.
$
On the other hand, relations
(\ref{LS00}) can be evidently rewritten as
\beaa
&&
\p_{t_1}\Phi= -((F\Phi)\Phi^{-1})_-\Phi,
\nn
\\
&&
\p_{t_2}\Phi= -((G\Phi)\Phi^{-1})_-\Phi.
\eeaa
Then
\beaa
-\Delta=\left\{
\p_{t_1}((G\Phi)\Phi^{-1})_-
-
\p_{t_2}((F\Phi)\Phi^{-1})_-
+
[((G\Phi)\Phi^{-1})_-,((F\Phi)\Phi^{-1})_-]
\right\}\Phi,
\eeaa
and $\Delta \Phi^{-1}$ contains only negative powers
in $\lambda$,
$
(\Delta \Phi^{-1})_+=0,
$
thus $\Delta=0$.
\end{proof}
\subsection*{Matrix dressing on the background}
Function $\Phi$ 
for matrix extension of the basic hierarchy
(\ref{analyticity0}),(\ref{genSato})
can be constructed using a matrix 
Riemann-Hilbert (RH) problem
\bea
\Phi_{\text{in}}=\Phi_{\text{out}}R(\Psi^0,\dots,\Psi^N),
\label{RH1}
\eea
defined on some curve in the complex plane 
(we will usually consider the unit circle).
Here $\Psi^0,\dots,\Psi^N$ are 
wave functions of the basic
hierarchy taken on the curve.

We suggest that solution $\Phi$ of the RH problem
is bounded
and invertible, normalisation by unit matrix 
at infinity fixes the gauge
and leads to closed systems of equations 
for matrix coefficients
of extended Lax pairs. The dispersionless hierarchy 
corresponding
to integrable distribution with the basis $X_i$ 
plays the role of
the background.

For canonically normalised matrix-valued function $\Phi$ 
the expansion at infinity is of the form (\ref{Phiser0})
\beaa
\Phi_\text{out}=
\text{I}+\sum_{n=1}^\infty \Phi_n(\mathbf{t})\lambda^{-n}
\eeaa
Sometimes it is useful to retain gauge freedom 
and consider solutions of 
the RH problem with the expansion
\be
\wt\Phi=
\wt\Phi_0
+\sum_{n=1}^\infty \wt\Phi_n(\mathbf{t})\lambda^{-n},
\label{Phiser1}
\ee
canonically normalised function is then expressed as
$\Phi={\wt\Phi_0}^{-1}\wt\Phi$.

From matrix
RH problem we get analyticity 
of the functions $(X_i\Phi)\Phi^{-1}$ and
of the
matrix-valued form
\beaa
\Omega=\Omega_0\wedge\d\Phi\cdot \Phi^{-1},
\eeaa
leading to additional generating relation for 
the extended hierarchy
\bea
(\Omega_0\wedge\d\Phi\cdot \Phi^{-1})_-=0.
\label{genext}
\eea 
Together generating relations 
(\ref{analyticity0}), 
(\ref{genext})
imply 
polinomiality 
of the series $(X_i\Phi)\Phi^{-1}$ and
Lax-Sato equations for the series 
(\ref{Phiser0}), (\ref{form0}), (\ref{formk})
defining the evolution of these series.

For linearly degenerate case $\Psi^0=\l$ 
vector fields do not contain the derivative
over $\lambda$, the dressing data contain 
an explicit dependence over the curve parameter,
for which it is enough to be continuous,
and to construct solutions we can also use
the $\dbar$ problem \cite{BK05}
\beaa
\dbar\Phi=\Phi 
R(\lambda,\bar\lambda,\Psi^1,\dots,\Psi^N),
\eeaa
that expands the opportunities to construct
explicit solutions considerably.
\subsection*{Lax-Sato equations and extended vector fields}
The second generating relation gives Lax-Sato
equations for series $\Phi$ on the vector field background
\bea
&&
\partial^k_n {\Psi}=V^k_n(\lambda)\Psi,
\label{LS1}
\\
&&
\partial^k_n {\Phi}=\left(V^k_n(\lambda)-
((V^k_n(\lambda)\Phi)\cdot\Phi^{-1})_+\right)\Phi,
\label{LS2}
\eea
where vector fields $V^k_n(\lambda)$ are defined 
by formula (\ref{genSato}).
Respectively, extended vector fields  are expressed as
\bea
\nabla^k_n=
\partial^k_n -V^k_n(\lambda)
+(V^k_n(\lambda)\Phi)\cdot\Phi^{-1})_+,
\eea
in gauge-invariant form
\bea
\wt\nabla^k_n=
\partial^k_n -V^k_n(\lambda)
-((\partial^k_n -V^k_n(\lambda))\wt\Phi)\cdot\wt\Phi^{-1})_+.
\eea
Lax-Sato equations for the first flows are given by
formulae (\ref{genlinear}) and (\ref{LS2}),
\bea
\partial^k_1\Phi
=
\left(
\lambda \partial_k
- (\partial_k\hat u)
-(\partial_k u_0)\partial_\lambda -  (\partial_k \Phi_1)
\right)\Phi,
\quad 0<k\leqslant N,
\label{genlinear1}
\eea
where $\hat u$ is a vector field, 
$\hat u=\sum_{p=1}^N u_p \p_p$ 
(see also  (\ref{Gensystem})),
and corresponding extended vector fields are
\bea
\nabla^k_1
=\partial^k_1 
-\lambda \partial_k
+ (\partial_k\hat u)
+ (\partial_k u_0)\partial_\lambda
+ (\partial_k \Phi_1),
\label{genlinear2}
\eea
in gauge-invariant form
\bea
&&
\nabla^k_1
=\partial^k_1 
-\lambda \partial_k
+ (\partial_k\hat u)
+ (\partial_k u_0)\partial_\lambda
+ \lambda A_k + B_k,
\label{gi}
\\
&&\qquad
A_k=(\partial_k \wt\Phi_0){\wt\Phi_0}^{-1},
\nn\\
&&\qquad
B_k=(\partial_k \wt\Phi_1){\wt\Phi_0}^{-1}
-(\partial_k \wt\Phi_0){\wt\Phi_0}^{-1}
\wt\Phi_1{\wt\Phi_0}^{-1}
-((\partial^k_1 + (\partial_k\hat u)) \wt\Phi_0)
{\wt\Phi_0}^{-1}.
\nn
\eea
Substituting series for $\Phi$ (\ref{Phiser0})
to  Lax-Sato equations (\ref{genlinear1}), 
we get recursion relations
expressing all the coefficients of the series 
through the first coefficient $\Phi_1$
and scalar functions $u_0,\dots u_n$
\beaa
\partial_k \Phi_{m+1}=
(\partial^k_1 
+ (\partial_k\hat u))\Phi_m
-(m-1) (\partial_k u_0)\Phi_{m-1}
+ (\partial_k \Phi_1)\Phi_m
\eeaa
Commutation relation for each pair of extended vector
fields (\ref{genlinear2}) gives $(N+2)$-dimensional
dispersionless equation (\ref{Gensystem}) and a matrix
$(N+2)$-dimensional equation of SDYM type on the background
of its solution,
\bea
(\partial^k_1+ (\partial_k\hat u))\partial_q \Phi_1
-
(\partial^q_1+ (\partial_q\hat u))\partial_k \Phi_1
+[(\partial_k \Phi_1),(\partial_q \Phi_1)]=0
\label{eqN}
\eea
In gauge invariant form
\beaa
&&[\nabla_k,\nabla_q]=0,
\\
&&[\nabla^k_1 , \nabla^q_1]
=(\partial_k u_0)\nabla_q
-(\partial_q u_0)\nabla_k
\\
&&
[\nabla^k_1, \nabla_q]=[\nabla^q_1, \nabla_k],
\\
\text{where}
&&
\nabla^p_1=\partial^p_1+ (\partial_p\hat u) + B_p,
\quad \nabla_p = \partial_p  + A_p, \quad
1\leqslant p\leqslant N,
\eeaa
we have three equations for four matrix functions
$A_k,A_q$, $B_k,B_q$
on the background 
of the basic dispersionless system.

For the case of trivial background $\hat u=0$
we get a set of consistent  4-dimensional
SDYM equations in a space of $2N$ variables
\beaa
\partial^k_1 \partial_q \Phi_1
-
\partial^q_1 \partial_k \Phi_1
+[(\partial_k \Phi_1),(\partial_q \Phi_1)]=0,
\eeaa
in gauge invariant form
\beaa
&&[\nabla_k,\nabla_q]=0,
\\
&&[\nabla^k_1 , \nabla^q_1]
=0
\\
&&
[\nabla^k_1, \nabla_q]=[\nabla^q_1, \nabla_k],
\\
\text{where}
&&
\nabla^p_1=\partial^p_1 + B_p,
\quad \nabla_p = \partial_p  + A_p, \quad
1\leqslant p\leqslant N.
\eeaa

The case $N=2$ was considered in the work
\cite{LVB17}, where it was demonstrated that 
equations on the background 
have a very direct and natural geometric interpretation.
For this case we have two extended 
vector fields (\ref{genlinear2}),
\beaa
\nabla^1_1
=\partial^1_1 
-\lambda \partial_1
+ (\partial_1 u_1)\p_1
+ (\partial_1 u_2)\p_2
+ (\partial_1 u_0)\partial_\lambda
+ (\partial_1 \Phi_1),
\\
\nabla^2_1
=\partial^2_1 
-\lambda \partial_2
+ (\partial_2 u_1)\p_1
+ (\partial_2 u_2)\p_2
+ (\partial_2 u_0)\partial_\lambda
+ (\partial_2 \Phi_1).
\eeaa
giving a Lax pair for 4-dimensional
equation with independent variables
$t^1_1$, $t^2_1$, $x_1$, $x_2$,
\bea
(\partial^1_1+ (\partial_1\hat u))\partial_2 \Phi_1
-
(\partial^2_1+ (\partial_2\hat u))\partial_1 \Phi_1
+[(\partial_1 \Phi_1),(\partial_2 \Phi_1)]=0
\label{eqN4}
\eea
representing
a general local form of SDYM equations on the
background of self-dual conformal structure \cite{DFK15}
(for neutral signature, modulo coordinate transformations
and a gauge). In  terms of linear operator 
$Q$ (\ref{Q}) representing a conformal structure in 
the form of symmetric bivector (inverse metric)
this equation reads
\beaa
Q\Phi_1=
[(\partial_1 \Phi_1),(\partial_2 \Phi_1)]
\eeaa
For second heavenly equation (\ref{Heav}),
describing general self-dual vacuum solutions of Einstein
equations
(for real case with neutral signature), we have
\beaa
\left(
\partial_w\p_x
+\partial_z\p_y 
+\Theta_{yy}\p_x\p_x
+\Theta_{xx}\p_y\p_y 
-2\Theta_{xy}\p_x\p_y 
\right)
\Phi_1=
[(\partial_1 \Phi_1),(\partial_2 \Phi_1)],
\eeaa
the metric of the background space is given by 
expression
(\ref{ASDmetric1}),
\beaa
g=dwdx+dzdy-\Theta_{yy} dw^2 + 2\Theta_{xy}dwdz  
- \Theta_{xx} dz^2.
\eeaa


It is hard to expect such a simple geometric 
interpretation in the case $N>2$. First, there is no
geometric interpretation of general dispersionless
equations (\ref{Gensystem}), it is only clear 
that they correspond to some 
special set of commutation relations
of $2N$ vector fields in $2N$-dimensional space (if we
consider all pairs of indices).
However, a reduced multidimensional
linearly-degenerate case with 
Hamiltonian vector fields corresponds to hyper-K\"ahler
hierarchies \cite{Takasaki89}, which are of geometric origin.
This could probably be a good starting point for geometric
interpretation of multidimensional matrix equations on 
dispersionless background.
In the hyper-K\"ahler case the 
dimension is even,
N=2M,
covariant vector fields (\ref{genlinear})
are of the form
\bea
\nabla^k_1
=\partial^k_1 
-\lambda \partial_k
+\{(\p_k\Theta),\dots\}_{\mathbf{x}}
+ (\partial_k \Phi_1),
\label{genlinear21}
\eea
where
\beaa 
\{f,\dots\}_{\mathbf{x}}=
\sum_{p=1}^{M}(\p_{M+p} f)\p_p
- (\p_{p} f)\p_{M+p}, 
\\
\{f, g\}_{\mathbf{x}}=
\sum_{p=1}^{M}(\p_{M+p} f)(\p_p g)
- (\p_{p} f)(\p_{M+p} g).
\eeaa
Equations (\ref{Gensystem})
reduce to the first equations of
hyper-K\"ahler hierarchy
\beaa
&&
\partial^k_1\p_q\Theta
-\partial^q_1\p_k\Theta
+\{\p_k \Theta,\p_q\Theta\}_{\mathbf{x}}=0,
\eeaa
for M=1 we obtain second heavenly equation
(\ref{Heav}).
Matrix equations (\ref{eqN})
on the background of hyper-K\"ahler hierarchy
take the form
\beaa
\partial^k_1\partial_q \Phi_1
-
\partial^q_1\partial_k \Phi_1
+
\{\partial_k\Theta, \partial_q \Phi_1\}_\mathbf{x}
-
\{\partial_q\Theta, \partial_k \Phi_1\}_\mathbf{x}
=
[(\partial_q \Phi_1),(\partial_k \Phi_1)].
\eeaa

\subsection*{Abelian case and Penrose formula}
In the Abelian case equations 
(\ref{eqN}) become linear,
l.h.s. represents an action of 
linear differential operators of the second order 
on the scalar
function $\phi_1$ (we will use $\phi$ instead of 
$\Phi$ in Abelian case). But nevertheless 
these equations could
be of interest for several reasons. 
First, in three and
four dimension, where we have an interpretation 
of equations in terms of gauge fields, 
the Abelian case corresponds to electromagnetic fields on
geometric background and could be 
of interest by itself. Second, the arising 
linear operators are connected with the 
symbol of linearisation of basic dispersionless equations 
and can be useful for the study of stability of solutions
and singularities of these equations.
For example, for second heavenly equation (\ref{Heav})
linear operator $Q$ is exactly the linearisation
of the equation.
And finally, the scalar RH problem (\ref{RH1}) 
can be solved
explicitly, and as a result we obtain an analogue
of Penrose formula in curved space. If we know some 
solution of dispersionless system together with a 
general wave function on the unit circle, this formula
gives a solution of corresponding 
linear equations on the background, depending
on an arbitrary function of (N+1) variables.

Thus in the Abelian case we have a scalar series
(\ref{Phiser0})
\bea
\phi=
1+\sum_{n=1}^\infty \phi_n(\mathbf{t})\lambda^{-n},
\label{Phiser01}
\eea
Lax-Sato equations for the first flows are
\bea
\partial^k_1\phi
=
\left(
\lambda \partial_k
- (\partial_k\hat u)
-(\partial_k u_0)\partial_\lambda -  (\partial_k \phi_1)
\right)\phi,
\quad 0<k\leqslant N,
\label{genlinear11}
\eea
and equations (\ref{eqN}) become linear,
\bea
\left((\partial^k_1+ (\partial_k\hat u))\partial_q 
-
(\partial^q_1+ (\partial_q\hat u))\partial_k 
\right)\phi_1
=0.
\label{eqN1}
\eea
A special solutions to Lax-Sato equations (\ref{genlinear11})
(and the whole set 
of Abelian Lax-Sato equations (\ref{LS2}))
is given by the Jacobian $J_0$ (\ref{J_0}) (see
\cite{LVB11} for more detail),
\beaa
\partial^k_1 J_0
=
\left(
\lambda \partial_k
- (\partial_k\hat u)
-(\partial_k u_0)\partial_\lambda -  
(\partial_k \sum_{i=1}^N \p_i u_i))
\right) J_0,
\quad 0<k\leqslant N,
\eeaa
and the function $\phi_1=\sum_{i=1}^N \p_i u_i$
(divergence of vector field $\hat u$) is a special
solution to equations (\ref{eqN1}).

Riemann-Hilbert problem (\ref{RH1}) in the
scalar case
\beaa
\phi_{\text{in}}=\phi_{\text{out}}R(\Psi^0,\dots,\Psi^N),
\eeaa
is solved in the standard way after
taking logarithm
\beaa
\ln \phi_{\text{in}} - \ln \phi_{\text{out}}
=r (\Psi^0,\dots,\Psi^N), \quad R=\e^r,
\eeaa
by the explicit formula
\beaa
\ln \phi=
\frac{1}{2\pi i}\oint 
\frac{r (\Psi^0,\dots,\Psi^N)}{\mu-\lambda} d\mu.
\eeaa
Respectively,
\beaa
\phi=
\exp\left(
\frac{1}{2\pi i}\oint 
\frac{r (\Psi^0,\dots,\Psi^N)}{\mu-\lambda} d\mu
\right),
\eeaa
and for the function $\phi_1$ we have
\bea
\phi_1=-
\frac{1}{2\pi i}\oint 
r (\Psi^0,\dots,\Psi^N) d\mu
\label{Psi0}
\eea
Here $r (\Psi^0,\dots,\Psi^N)$ is an arbitrary
complex analytic function of its arguments,
it represents a general solution (wave function)
of linear equations
(\ref{genlinear11}), 
defined on the unit circle in the complex plane.
Any wave function $\Psi(\bbox{t},\lambda)$
defined on the unit circle (or other
closed curve) gives a solution to equations
(\ref{eqN1}),
\bea
\phi_1=-
\frac{1}{2\pi i}\oint 
\Psi(\bbox{t},\mu) d\mu,
\label{Psi}
\eea
it can be easily checked directly.
Indeed,
\beaa
\left((\partial^k_1+ (\partial_k\hat u))\partial_q 
-
(\partial^q_1+ (\partial_q\hat u))\partial_k 
\right)\Psi
=((\p_q u_0)\p_k - (\p_k u_0)\p_q)\p_\l \Psi,
\eeaa
and integration by $\l$ over a closed contour
gives zero, thus expression (\ref{Psi}) is
a solution to equations (\ref{eqN1}).

For N=2 we have one linear equation (\ref{eqN1})
\bea
&&
Q\phi_1=0,
\label{Qq2}
\\
&&
Q=
\partial_w\p_x
-\partial_z\p_y 
+(\p_y u_1)\p_x\p_x 
-(\p_x u_2)\p_y\p_y 
-((\p_x u_1)-(\p_y u_2))\p_x\p_y,
\nn
\eea
where operator $Q$ (\ref{Q}) corresponds
to symmetric bivector of self-dual
conformal structure; notations for times 
correspond to equation (\ref{Dun}).
This equation represent a general 
local form of Abelian self-dual Yang-Mills equations
(self-dual equations of electromagnetic field)
on the background of conformal
structure with neutral signature in a special
gauge.
The corresponding metric (representative of conformal
structure) is given by formula (\ref{ASDmetric0}),
\beaa
&&
g=dw dx
-dz dy
-(\p_y u_1) dw dw
-((\p_x u_1)-(\p_y u_2)) dz dw 
+ (\p_x u_2)dz dz.
\eeaa

Solutions to equation (\ref{Qq2})
are given by formula (\ref{Psi0})
with N=2,
\bea
\phi_1=-
\frac{1}{2\pi i}\oint 
r (\Psi^0,\Psi^1,\Psi^2) d\mu.
\label{Psi01}
\eea
For trivial background $\hat u=0$,
$\Psi^0=\lambda$, 
$\Psi^1=\sum_{n=0}^\infty t^1_n \l^{n}$,
$\Psi^2=\sum_{n=0}^\infty t^2_n \l^{n}$,
equation (\ref{Qq2}) reduces
to 4-dimensional wave equation for neutral
signature with constant coefficients
\beaa
\left(
\partial_w\p_x
-\partial_z\p_y 
\right)\phi_1=0,
\eeaa
formula (\ref{Psi01}) takes the form
\beaa
\phi_1=-
\frac{1}{2\pi i}\oint 
r (\l, \l z +x,\l w + y) d\mu.
\eeaa
This formula is easily recognised as a version
of Penrose formula \cite{Pen69}, \cite{Dun}
for solutions of the wave equation
written for the case of neutral signature.
Thus formula (\ref{Psi01})
can be considered as a generalisation 
of Penrose formula to the case of operators
connected with the self-dual conformal structure.

\subsection*{Acknowledgements} The author is
grateful to E.A. Kusnetsov for useful comment
concerning the topic of this work.
\subsection*{Conflict of Interest} 
The author declares no conflicts of interest.

\end{document}